\documentclass[prd,floats,aps,epsf,twocolumn,showpacs]{revtex4-1}

\usepackage{amsmath}
\usepackage{amssymb}
\usepackage{bm}
\usepackage{graphicx}
\usepackage{epstopdf}

\begin{document}

\title{An integral equation for distorted wave amplitudes}

\author{Luca Visinelli}
\email{visinelli@utah.edu}
\author{Paolo Gondolo}
\email{paolo@physics.utah.edu}
\affiliation{Department of Physics and Astronomy, University of Utah, 115 South 1400 East \#201, Salt Lake City, Utah 84112-0830, USA}

\date{\today}

\begin{abstract} \noindent
We derive a new integral equation that allows the calculation of the scattering or annihilation amplitude of two particles subjected to two potentials when the corresponding amplitude for one potential only is known. We assume that scattering or annihilation occurs through one of the potentials, while the other potential affects the particle wave functions. Our expression is valid for any choice of the distorting potential and for any particle model. Our technique does not require the expansion of the amplitude into partial waves, and allows the study of models that are generally difficult to solve by means of the Schroedinger equation.
\end{abstract}

\pacs{03.65.Nk, 11.10.St}

\maketitle

There are many examples in nature in which a physical system is affected by two distinct potentials. For example, nuclear beta decay occurs through weak interactions, but electromagnetic forces between the charged nucleus and the emitted electron affect the motion of the final particles by decelerating them \cite{fermi,landau1}. Another example is nucleon-nucleon scattering \cite{bethe_salpeter}, in which the exchange of virtual mesons before the scattering can be described by an effective potential that enhances the amplitude for the process. A similar situation has been invoked in dark matter studies to generate anomalously high annihilation cross sections at low velocities in relation to recent cosmic ray data (an effect dubbed the Sommerfeld enhancement after Ref.~\cite{sommerfeld}). Another example is provided by the quark-antiquark bound state system \cite{delbourgo}, in which there is a potential for quark confinement and another for quark decay. Still another example occurs in the minimal supersymmetric standard model when the lightest and second lightest supersymmetric particles are close in mass and convert into each other prior to their final annihilation \cite{hisano}.

In all of these examples, one potential (from here on indicated with $\hat{W}$) is responsible for the final scattering, annihilation or decay, while another potential (here referred to as the distorting potential $\hat{V}$) affects the initial particle wave functions and thus modifies the transition amplitude. 

In this Letter, we derive an exact integral equation for the transition amplitude in the presence of two potentials $\hat V$ and $\hat W$ in terms of the transition amplitude with only one of the potentials, say $\hat W$. We also derive the equation relating the corresponding initial-state wave functions for two-particle states. To the extent of our knowledge, these equations have not been presented before in the literature.

In our derivation, we do not make any assumption regarding the strength of one potential compared to the other. In other words, we do not invoke the so-called Distorted Wave Born Approximation (DWBA) \cite{davydov, weinberg}, although we discuss the simplification of our integral equation when the DWBA is applied. 

Although our derivation is general, we present it here for the common case of two incident particles. The final state is completely arbitrary. Consider a process in which two particles $\chi_1$ and $\chi_2$, of masses $m_1$ and $m_2$, respectively, scatter or annihilate into a multi-particle final state,
\begin{equation} \label{process}
\chi_1 + \chi_2 \to \chi'_1 + ... + \chi'_n.
\end{equation}
The particles that take part in the process can be fermions or bosons. In the center-of-mass (CM) frame, $\chi_1$ has momentum ${\bf p_1} \equiv {\bf p}$, and $\chi_2$ has momentum ${\bf p_2} \equiv -{\bf p}$. If one or both particles have spin, we quantize the total spin of the initial two-particle state along the axis defined by the direction of ${\bf p}_1$. Thus, if $\lambda_1$ and $\lambda_2$ denote the helicities of particles 1 and 2, i.e.\ their spin projection in the direction of their respective momenta, the total CM spin of the initial state has projection $\Lambda_{\bf p}=\lambda_1-\lambda_2$ along the direction of ${\bf p}$. 

In labeling the initial two-particle state we need to distinguish between plane-wave states, scattering waves in the presence of the potential $\hat W$ alone, and scattering waves in the presence of both potentials $\hat V$ and $\hat W$. We denote these states by
$|{\bf p}S\Lambda_{\bf p}\eta\rangle$, $|\psi^{W\pm}_{{\bf p}S\Lambda_{\bf p}\eta}\rangle$, and $|\psi^{\pm}_{{\bf p}S\Lambda_{\bf p}\eta}\rangle$, respectively. Here $S$ is the total CM spin, the index $\pm$ distinguishes between incoming ($-$) and outgoing ($+$) scattering waves,  and $\eta$ (later suppressed) denotes additional quantum numbers besides momentum and spin. We also use a condensed notation $\alpha\equiv({\bf p},S,\Lambda_{\bf p},\eta)$ for the ensemble of quantum numbers.

We remark at this point that for the process in Eq.~(\ref{process}) with $\chi_1$ and $\chi_2$ in the initial state, one should use outgoing scattering waves since it is these waves that reduce to plane waves in the infinite past ($t\to-\infty$). For this reason, in the following we restrict ourselves to outgoing ($+$) waves.

The states $|\psi^{W+}_\alpha\rangle$ satisfy the Schroedinger equation
\begin{equation} \label{schroedinger}
(\hat{H}_0 + \hat{W})\,|\psi^{W+}_\alpha\rangle = E_\alpha\,|\psi^{W+}_\alpha\rangle.
\end{equation}
Here, $\hat{H}_0$ is the free hamiltonian, and according to the prescription in scattering theory, $E_\alpha$ is the total energy of the two particles in the CM frame, equal to the total energy in the absence of the potential: 
\begin{equation}
\label{free_hamiltonian}
\hat{H}_0|\alpha\rangle=E_\alpha|\alpha\rangle.
\end{equation}
In the non-relativistic limit, 
$E_\alpha = {\bf p}^2/(2\mu)$,
where
$\mu=m_1m_2/(m_1+m_2)$
is the reduced mass of the initial particles.

The states $|\psi^{W+}_\alpha\rangle$ satisfy the completeness relation
\begin{equation}\label{completeness}
\int d\alpha\, |\psi^{W+}_\alpha\rangle\langle \psi^{W+}_\alpha | = 1,
\end{equation}
for an appropriate integration measure that can contain a sum over discrete quantum numbers. The states  $|\psi^{W+}_\alpha\rangle$ are normalized so that
\begin{equation} \label{normalization}
\langle \psi^{W+}_{\alpha}|\,\psi^{W+}_{\alpha'}\rangle = (2\pi)^3\delta(\alpha - \alpha'),
\end{equation}
where similarly to the integration measure, the Dirac delta function can include Kronecker deltas in the case of discrete eigenvalues. We remark that the symbols $\alpha$ and $\alpha'$ appearing in Eqs.~(\ref{completeness}) and~(\ref{normalization}) contain both continuum and bound states.

We now add the distorting potential $\hat{V}$. The states $|\psi^+_\alpha\rangle$ satisfy the equation
\begin{equation} \label{schroedinger1}
(\hat{H}_0 + \hat{W}+\hat{V})\,|\psi^+_\alpha\rangle = E_\alpha\,|\psi^+_\alpha\rangle,
\end{equation}
where, according to scattering theory, $E_\alpha$ is the same energy appearing in Eq.~(\ref{schroedinger}). Following standard procedure (see, e.g., \cite{weinberg,davydov}),
Eq.~(\ref{schroedinger1}) can be cast into the Lippmann-Schwinger form
\begin{equation}\label{LS}
|\psi^+_\alpha\rangle = |\psi^{W+}_\alpha\rangle + (E_\alpha - \hat{H}_0 - \hat{W} + i\epsilon)^{-1}\,\hat{V}\,|\psi^+_\alpha\rangle.
\end{equation}
To see that Eq.~(\ref{schroedinger1}) and Eq.~(\ref{LS}) are identical, one can multiply Eq.~(\ref{LS}) on the left by the operator $E_\alpha - \hat{H}_0 - \hat{W}$, after which the term $(E_\alpha - \hat{H}_0 - \hat{W})|\psi^{W+}_\alpha\rangle$ vanishes because of Eq.~(\ref{schroedinger}). The extra term $+i\epsilon$ in Eq.~(\ref{LS}) is introduced in order to deal with the singularity in the operator $(E_\alpha - \hat{H}_0 - \hat{W})^{-1}$, and takes into account the sign (+) describing the out-going state $|\psi^+_\alpha\rangle$.

{}From Eq.~(\ref{LS}) we derive our main result as follows. We insert a completeness relation~(\ref{completeness}) into the second term on the right hand side of Eq.~(\ref{LS}), thus
\begin{align}
& (E_\alpha - \hat{H}_0 - \hat{W} + i\epsilon)^{-1}\,\hat{V}\,|\psi^+_\alpha\rangle = 
\nonumber \\
& = \int d\gamma\,(E_\alpha - \hat{H}_0 - \hat{W} + i\epsilon)^{-1}\,|\psi^{W+}_\gamma\rangle\langle\psi^{W+}_\gamma|\, \hat{V}\,|\psi^+_\alpha\rangle=
\nonumber \\
& = \int d\gamma\frac{|\psi^{W+}_\gamma\rangle\,\langle\psi^{W+}_\gamma|\, \hat{V}\,|\psi^+_\alpha\rangle}{E_\alpha -E_\gamma+ i\epsilon}.
\end{align}
The label $\gamma$ describes the set of quantum numbers associated with the intermediate state $|\psi^{W+}_\gamma\rangle$. Defining 
\begin{equation}
T_{\gamma\alpha} = \langle\psi^{W+}_\gamma|\, \hat{V}\,|\psi^{+}_\alpha\rangle,
\end{equation}
Eq.~(\ref{LS}) becomes
\begin{equation} \label{LS2}
|\psi^+_\alpha\rangle = |\psi^{W+}_\alpha\rangle +  \int d\gamma\frac{ |\psi^{W+}_\gamma\rangle\,T_{\gamma\alpha}}{E_\alpha -E_\gamma+ i\epsilon}.
\end{equation}

The quantity $T_{\gamma\alpha}$, which coincides with the difference of the $T$-matrices for the process in Eq.~(\ref{process}) when both $\hat{V}$ and $\hat{W}$ or only $\hat{W}$ are present, satisfies the Lippmann-Schwinger equation 
\begin{equation} \label{transition3D}
T_{\gamma\alpha} = V^W_{\gamma\alpha} + \int d\gamma'\,\frac{V^W_{\gamma\gamma'}}{E_{\alpha} - E_{\gamma'} + i\epsilon}\,T_{\gamma'\alpha},
\end{equation}
where
\begin{equation} \label{potential}
V^W_{\gamma\gamma'} \equiv \langle\psi^{W+}_\gamma|\hat{V}|\psi^{W+}_{\gamma'}\rangle
\end{equation}
are the matrix elements of $\hat{V}$ in the basis $|\psi^{W+}_{\alpha}\rangle$.
Eq.~(\ref{transition3D}) can be obtained by multiplying Eq.~(\ref{LS}) on the left by $\langle\psi^{W+}_\gamma|\, \hat{V}$, and inserting a completeness relation in a way similar to the derivation of Eq.~(\ref{LS2}). 

We now use Eq.~(\ref{transition3D}) to eliminate $T_{\gamma\alpha}$ from Eq.~(\ref{LS2}). We expand Eq.~(\ref{transition3D}) recursively in powers of $V^{W}$,
\begin{align}
&T_{\gamma\alpha} = V^W_{\gamma\alpha} + \int d\gamma'\frac{V^W_{\gamma\gamma'}\,V^W_{\gamma'\alpha}}{E_{\alpha} - E_{\gamma'} + i\epsilon} +
\nonumber \\
&+\int d\gamma'd\gamma'' \frac{V^W_{\gamma\gamma'}\,V^W_{\gamma'\gamma''}V^W_{\gamma''\alpha}}{(E_{\alpha} - E_{\gamma'} + i\epsilon)(E_{\alpha} - E_{\gamma''} + i\epsilon)} + \cdots ,
\end{align}
and insert this expansion into Eq.~(\ref{LS2}),
\begin{align}
& |\psi^+_\alpha\rangle = |\psi^{W+}_\alpha\rangle + \int d\gamma \frac{|\psi^{W+}_\gamma\rangle\,V^W_{\gamma\alpha}}{E_{\alpha} - E_{\gamma} \pm i\epsilon} + 
\nonumber \\
\label{transition1}
&+\int d\gamma d\gamma' \frac{|\psi^{W+}_\gamma\rangle\,V^W_{\gamma\gamma'}\,V^W_{\gamma'\alpha}}{(E_{\alpha} - E_{\gamma} + i\epsilon)(E_{\alpha} - E_{\gamma'} + i\epsilon)} + \cdots .
\end{align}
We finally notice that the last expression corresponds to the series expansion of
\begin{equation} \label{master_equation1}
|\psi^+_\alpha\rangle =|\psi^{W+}_\alpha\rangle + \int d\gamma \frac{|\psi^+_\gamma\rangle\,V^{W}_{\gamma\alpha}}{E_{\alpha} - E_{\gamma} + i\epsilon}.
\end{equation}
Eq.~(\ref{master_equation1}) is one of our two main results (the other one is Eq.~(\ref{master_equation2}) below). 

Eq.~(\ref{master_equation1})  relates the state $|\psi^+_\alpha\rangle$, solution of the Schroedinger equation in the presence of both potentials $\hat{V}$ and $\hat{W}$, to the state $|\psi^{W+}_\alpha\rangle$, solution of the Schroedinger equation in the presence of the potential $\hat{W}$ only. Eq.~(\ref{master_equation1})  can be used to find the relative normalization of the  $|\psi^+_\alpha\rangle$ and $|\psi^{W+}_\alpha\rangle$ states.

We stress that no assumption regarding the strength of $\hat{V}$ relative to $\hat{W}$ has been made in deriving Eq.~(\ref{master_equation1}).  The potential $\hat{W}$ does not enter Eq.~(\ref{master_equation1}) directly, so to find $|\psi^+_\alpha\rangle$ one only needs to know the states $|\psi^{W+}_\alpha\rangle$ and the potential $\hat{V}$. We remark that Eq.~(\ref{master_equation1}) cannot be obtained from Eq.~(\ref{LS}) by simply inserting a completeness relation in the way we derived Eqs.~(\ref{LS2}) and~(\ref{transition3D}). 

{}From Eq.~(\ref{master_equation1}) we derive an integral equation for the transition amplitude of process~(\ref{process}). Introducing the final state $|\phi_\beta\rangle$, where $\beta$ labels the quantum numbers of the final particles $\chi'_1,...,\chi'_n$, the transition amplitude for process~(\ref{process}) when only $\hat{W}$ is present is
\begin{equation}\label{bare_amplitude}
\mathcal{M}^0_{\beta\alpha} = \langle\phi_\beta|\psi^{W+}_\alpha\rangle.
\end{equation}
We refer to $\mathcal{M}^0_{\beta\alpha}$ as the {\it undistorted} amplitude, undistorted by the potential $\hat V$. When both $\hat{V}$ and $\hat{W}$ are present, the transition amplitude is
\begin{equation}\label{distorted_amplitude}
\mathcal{M}^+_{\beta\alpha} = \langle\phi_\beta|\psi^+_\alpha\rangle.
\end{equation}
We refer to $\mathcal{M}^+_{\beta\alpha}$ as the {\it distorted} amplitude. Multiplying Eq.~(\ref{master_equation1}) on the left by $\langle\phi_\beta|$, we obtain
\begin{equation} \label{master_equation2}
\mathcal{M}^+_{\beta\alpha} = \mathcal{M}^0_{\beta\alpha} + \int d\gamma \frac{\mathcal{M}^+_{\beta\gamma}\,V^W_{\gamma\alpha}}{E_\alpha - E_\gamma + i\epsilon}.
\end{equation}
This is our second main result. Eq.~(\ref{master_equation2}) is an integral equation for the distorted amplitude, which can be solved if the undistorted amplitude and $V^W_{\gamma\alpha}$ are known.

In Ref.~\cite{visinelli} we show that Eq.~(\ref{master_equation2}) is the non-relativistic limit of sums of ladder diagrams in quantum field theory. Here we only sketch the proof graphically in Fig.~1. In this case, $\hat V$ and $\hat W$ are effective potentials that may include radiative corrections to any number of loops.

The graphical derivation allows the understanding of which quantum field theory diagrams are included in our Eq.~(\ref{master_equation2}). They are ladder diagrams containing multiple powers of the effective potential $\hat{V}$, while the other effective potential $\hat{W}$, responsible for the final interaction, appears only once. Radiative corrections like bremsstrahlung of particles from the initial or intermediate legs can also be treated in our equation by including the bremsstrahlung particles in a modified final state. Notice in this regard that non-relativistic quantum mechanics does not allow for the number of particles in the final state to change, so an $n_1$-particle process with extra $n_2$ bremsstrahlung particles should be considered as an $n_1\!+\!n_2$-particle process.

\begin{figure}[t]
\centering
\includegraphics[width=0.49\textwidth]{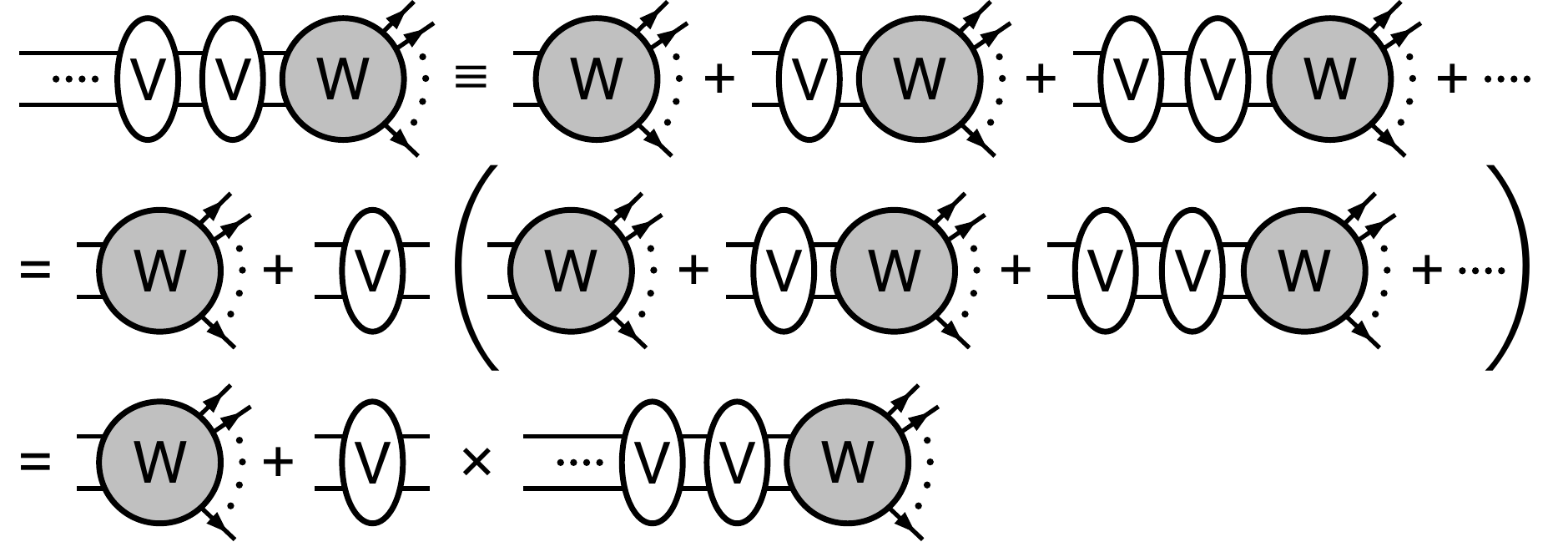}
\caption{Graphical derivation of the ladder equation that reduces to our Eq.~(\ref{master_equation2}) in the non-relativistic limit. Gray circles stand for the effective potential $\hat W$, including radiative corrections to any number of loops. White ovals indicate the effective potential $\hat{V}$ to any number of loops. The diagram on the LHS, indicating $\mathcal{M}^+_{\beta\alpha}$ [Eq.~(\ref{distorted_amplitude})], is the sum of an infinite ladder of effective potentials $\hat V$ and a final $\hat{W}$ interaction. The first diagram on the RHS stands for $\mathcal{M}^0_{\beta\alpha}$ [Eq.~(\ref{bare_amplitude})]. }
\label{Fig1}
\end{figure}

So far our derivation of Eqs.~(\ref{master_equation1}) and~(\ref{master_equation2}) has been exact. We now discuss Eq.~(\ref{master_equation2}) in the Born approximation with respect to the potential $\hat{W}$. In the first Born approximation, 
\begin{equation}
|\psi^{W+}_\alpha\rangle = |\alpha\rangle + O(\hat{W}),
\end{equation}
where $|\alpha\rangle$ is a free state of Eq.~(\ref{free_hamiltonian}).
To the same order in $\hat{W}$, the matrix elements in Eq.~(\ref{potential}) become
\begin{equation}
V^W_{\gamma\alpha} = \langle\gamma|\hat{V}|\alpha\rangle  + O(\hat{W}).
\end{equation}
Thus the distorted amplitude $\mathcal{M}^+_{\beta\alpha}$ in the first Born approximation in $\hat{W}$, which is known as the DWBA amplitude, satisfies the integral equation
\begin{equation} \label{master_equation3}
\mathcal{M}^+_{\beta\alpha} = \mathcal{M}^0_{\beta\alpha} + \int d\gamma \frac{\mathcal{M}^+_{\beta\gamma}\,V_{\gamma\alpha}}{E_\alpha - E_\gamma + i\epsilon}.
\end{equation}
The only difference from Eq.~(\ref{master_equation2}), besides the Born approximation implicit in $\mathcal M^0_{\beta\alpha}$, is the presence of $V_{\gamma\alpha} = \langle\gamma|\hat{V}|\alpha\rangle$ instead of $V^W_{\gamma\alpha}=\langle\psi^{W+}_\gamma|\hat{V}|\psi^{W+}_{\gamma'}\rangle$, i.e.\ plane wave states replace $\hat{W}$ scattering states in the matrix elements of $\hat{V}$. Eq.~(\ref{master_equation3}) is a good approximation whenever the potential $\hat W$ is weak, which is a common case.

For potentials $\hat V$ that do not depend on the spin of the particles, Eq.~(\ref{master_equation3}) with the Born approximation in $\hat{W}$ splits into separate independent equations for each value of the total spin $S$. (If $\hat W$ is also spin-independent, then an analogous derivation shows that the full Eq.~(\ref{master_equation2}) separates into a set of independent equations for each total spin $S$.) To see this, we factorize the two-particle free state $|\alpha\rangle=|{\bf p}S\Lambda_{\bf  p}\eta\rangle$ into an orbital part $|{\bf p}\eta\rangle$ and a spin part $|S\Lambda_{\bf p}\rangle$. Thus,
\begin{equation}
 |\alpha\rangle \equiv |{\bf p}S\Lambda_{\bf p}\eta\rangle = |{\bf p}\eta\rangle \, |S\Lambda_{\bf p}\rangle .
 \label{eq:factor1}
\end{equation}
Similarly, we factorize the intermediate state into
\begin{equation}
|\gamma\rangle \equiv |{\bf k}S'\Lambda'_{\bf k}\eta'\rangle = |{\bf k}\eta'\rangle\, |S'\Lambda'_{\bf k}\rangle.
 \label{eq:factor2}
\end{equation}
Here ${\bf k}$ is the internal three-momentum.  Notice that for $S = 0$, only the orbital parts exist: $|\alpha\rangle = |{\bf p}\eta\rangle$ and $|\gamma\rangle = |{\bf k}\eta'\rangle$.

For simplicity, we define the $+z$ axis along the initial momentum, and specify the direction of the internal momentum ${\bf k}$ using spherical coordinates $(\theta_k,\phi_k)$. 
Then the helicity state $|S'\Lambda'_{\bf k}\rangle$ with spin polarization along ${\bf k}$ can be obtained through a rotation from the helicity state $|S'\Lambda'_{\bf p}\rangle$ polarized along ${\bf p}$.
In the convention of Jacob and Wick (see Eq.~(6) in Ref.~\cite{JW}), the two helicity states are related through a rotation of Euler angles $(\phi_k,\theta_k, -\phi_k)$,
\begin{equation}
|S\Lambda'_{\bf k}\rangle = \mathcal{R}(\phi_k,\theta_k, -\phi_k)\,|S\Lambda'_{\bf p}\rangle.
\end{equation}
The rotation operator $\mathcal{R}(\phi_k,\theta_k, -\phi_k)$ is related to the Wigner $D$-matrix $ D^{(S)}_{MM'}$ through
\begin{align}
&\langle S\Lambda_{\bf p}|S\Lambda'_{\bf k}\rangle  =\langle S\Lambda_{\bf p}|\mathcal{R}(\phi_k,\theta_k, -\phi_k)|S\Lambda'_{\bf p}\rangle = 
\nonumber \\
&
\label{D-matrix}
=  D^{(S)}_{\Lambda_{\bf p},\Lambda'_{\bf p}}(\phi_k,\theta_k,-\phi_k) =  e^{-i(\Lambda_{\bf p}-\Lambda'_{\bf p})\phi_k}\,d^{(S)}_{\Lambda_{\bf p},\Lambda'_{\bf p}}(\theta_k).
\end{align}
The function $d^{(S)}_{MM'}(\beta)$, which is a real function and is known in the literature as the small $d$-matrix, is defined using the generator  $\hat J_{y}$ of rotations about the $y$-axis between spin states polarized along the $z$ axis:
\begin{equation}
d^{(S)}_{MM'}(\beta) = \langle SM | e^{-i\hat J_{y}\beta/\hbar}|SM'\rangle.
\end{equation}

Using Eqs.~(\ref{eq:factor1}), (\ref{eq:factor2}), and the complex conjugate of Eq.~(\ref{D-matrix}) with $D^{(S)*}_{\Lambda_{\bf p},\Lambda'_{\bf p}}(\phi_k,\theta_k,-\phi_k) =D^{(S)}_{\Lambda_{\bf p},\Lambda'_{\bf p}}(-\phi_k,\theta_k,\phi_k) $, the matrix elements  $V_{\gamma\alpha}= \langle\gamma |\hat{V}|\alpha\rangle$ of a spin-independent potential factorize as
\begin{align}
V_{\gamma\alpha} & = \langle S'\Lambda'_{\bf k}|\langle{\bf k}|\hat{V}|{\bf p}\rangle |S \Lambda_{\bf p}\rangle 
= \langle S'\Lambda'_{\bf k}| S \Lambda_{\bf p}\rangle  \, \langle{\bf k}|\hat{V}|{\bf p}\rangle
\nonumber \\
& = \delta_{SS'} \, D^{(S)}_{\Lambda_{\bf p},\Lambda'_{\bf p}}(-\phi_k,\theta_k,\phi_k) \,\langle{\bf k}|\hat{V}|{\bf p}\rangle.
\end{align}
Here $ \langle{\bf k}|\hat{V}|{\bf p}\rangle $ is the Fourier transform of the distorting potential $V({\bf r})$,
\begin{equation}
\langle{\bf k}|\hat{V}|{\bf p}\rangle = \int d^3r\,V({\bf r})\,e^{-i({\bf k} - {\bf p})\cdot{\bf r}}
\equiv \tilde V({\bf p} -{\bf k}) .
\end{equation}
Hence
\begin{equation}
V_{\gamma\alpha}  = \delta_{SS'} D^{(S)}_{\Lambda_{\bf p},\Lambda'_{\bf p}}(-\phi_k,\theta_k,\phi_k) \,\tilde V({\bf p}-{\bf k}).
\label{eq:Vfactorize}
\end{equation}
Notice that for spin zero particles, one simply obtains $V_{\gamma\alpha}  = \tilde V({\bf p}-{\bf k})$. The Kronecker delta in Eq.~(\ref{eq:Vfactorize}) forbids any mixing of different total spins in Eq.~(\ref{master_equation3}). 

Eq.~(\ref{master_equation3}) then splits into a separate set of $2S+1$ equations for every value of the total spin $S$. To display the structure of the each set clearly, we introduce the abbreviated notation
\begin{equation}
\mathcal{M}^+_{S\Lambda}({\bf p}) \equiv \mathcal{M}^+_{\beta,{\bf p}S\Lambda_{\bf p}\eta},\quad
\mathcal{M}^0_{S\Lambda}({\bf p}) \equiv \mathcal{M}^0_{\beta,{\bf p}S\Lambda_{\bf p}\eta}.
\end{equation}
Then for $\Lambda=-S,-S+1,\ldots,S$, we find
\begin{align}
& \mathcal{M}^+_{S\Lambda}({\bf p}) =  \mathcal{M}^0_{S\Lambda}({\bf p}) + \nonumber \\
\label{master_equation4}
& \quad \sum_{\Lambda'=-S}^{S}\int \frac{d^3k}{(2\pi)^3} \frac{\mathcal{D}^{(S)}_{\Lambda\Lambda'}(-\phi_k,\theta_k,\phi_k)\,\tilde V({\bf p}-{\bf k})\,\mathcal{M}^+_{S\Lambda'}({\bf k})}{E_p - E_k + i\epsilon}.
\end{align}
In general, Eq.~(\ref{master_equation4}) has to be solved numerically for each value of $S$ to obtain $\mathcal{M}^+_{S\Lambda}({\bf p})$.

Eq.~(\ref{master_equation4}),  obtained in the DWBA, is valid for any spin of the annihilating particles. In particular, for bosonic particles or for particles which are in a singlet state, $S = \Lambda = 0$, $\mathcal{D}^{(0)}_{\Lambda\Lambda'}(-\phi_k,\theta_k,\phi_k) = 1$, and 
\begin{equation} \label{master_equation_singlet}
\mathcal{M}^+_{00}({\bf p}) = \mathcal{M}^0_{00}({\bf p}) +\int \frac{d^3k}{(2\pi)^3} \frac{\tilde V({\bf p}-{\bf k})\,\mathcal{M}^+_{00}({\bf k})}{E_p - E_k + i\epsilon}.
\end{equation}
This singlet equation, Eq.~(\ref{master_equation_singlet}), appears in Ref.~\cite{iengo} in a form limited to the specific case of a Yukawa potential $V({\bf r})$. Our equations apply to a generic potential and to any value of the total spin.

We have derived a new exact integral equation, Eq.~(\ref{master_equation2}), that gives the amplitude $\mathcal{M}^+_{\beta\alpha}$ for two particles to scatter or annihilate into a multi-particle state in the presence of two potentials $\hat{V}$ and $\hat{W}$ once the amplitude $\mathcal{M}^0_{\beta\alpha}$ for the same process in the absence of the potential $\hat{V}$ is known. We have also obtained a similar relation connecting the initial particle states in presence and absence of $\hat{V}$, see Eq.~(\ref{master_equation1}). We have also shown how our equation simplifies in the DWBA, Eq.(\ref{master_equation3}). For spin-independent potentials $\hat{V}$, our DWBA equation separates into independent sets, Eq.~(\ref{master_equation4}), one for each value of the total spin $S$ of the colliding particles. The same is true for our full equation~(\ref{master_equation2}) if also $\hat{W}$ is spin-independent.
Our integral equations~(\ref{master_equation2}) or~(\ref{master_equation4}) can be solved numerically for the amplitudes $\mathcal{M}^+_{\beta\alpha}$ or $\mathcal{M}^+_{S\Lambda}({\bf p})$, respectively, without an expansion in partial waves. Avoiding the partial wave expansion is convenient, for example, when the s-wave is slightly suppressed or there are p-wave resonances. Numerical studies of Eq.~(\ref{master_equation4}) will be presented in a separate paper \cite{visinelli}.

This work was supported in part by NSF award PHY-0456825 and NASA contract NNX09AT70G.

\end{document}